%% file: FANS.tex
\documentclass[sigconf]{acmart}

\AtBeginDocument{%
  \providecommand\BibTeX{{%
    \normalfont B\kern-0.5em{\scshape i\kern-0.25em b}\kern-0.8em\TeX}}}

\copyrightyear{2023}
\acmYear{2023}
\setcopyright{acmlicensed}\acmConference[WWW '23]{Proceedings of the ACM Web Conference 2023}{April 30-May 4, 2023}{Austin, TX, USA}
\acmBooktitle{Proceedings of the ACM Web Conference 2023 (WWW '23), April 30-May 4, 2023, Austin, TX, USA}
\acmPrice{15.00}
\acmDOI{10.1145/3543507.3583430}
\acmISBN{978-1-4503-9416-1/23/04}
  
\acmSubmissionID{6385}

\usepackage{hyperref}
\usepackage{natbib}
\usepackage{bm}
\usepackage{subcaption}

\usepackage{multirow}
\usepackage{color}
\usepackage{colortbl}
\usepackage{balance}
\newcommand{\ag}{\includegraphics{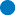}}
\newcommand{\nag}{\includegraphics{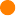}}
\newcommand{\tsc}{\includegraphics{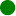}}

\newcommand{\cls}{\left \langle \texttt{CLS} \right \rangle}
\newcommand{\sep}{\left \langle \texttt{SEP} \right \rangle}
\newcommand{\mask}{\left \langle \texttt{MASK} \right \rangle}
\newcommand{\pad}{\left \langle \texttt{PAD} \right \rangle}
\newcommand{\front}{\mathbf{x}}

\newcommand{\rear}{\mathbf{y}}

\newcommand{\first}[1]{\textbf{#1}}
\newcommand{\second}[1]{\underline{#1}}

\newcommand{\nagbert}{{BERT4Rec$_\texttt{NAR}$ }}
\newcommand{\fansB}{{FANS$_\texttt{NAIVE}$ }}
\newcommand{\fansS}{{FANS$_\texttt{TSC-STEP}$ }}
\newcommand{\fansN}{{FANS$_\texttt{TSC-NAIVE}$ }}
\newcommand{\fansV}{{FANS$_\texttt{STEP}$ }}

\newcommand{\fansac}{\textcolor[HTML]{3E7A3E}}  
\newcommand{\fansuac}{\textcolor[HTML]{DB3636}}  

\begin{document}

\title[FANS: Fast Non-Autoregressive Sequence Generation for Item List Continuation]{FANS: Fast Non-Autoregressive Sequence Generation \\ for Item List Continuation}


\author{Qijiong Liu}
\email{jyonn.liu@connect.polyu.hk}
\affiliation{%
  \institution{The Hong Kong Polytechnic University}
  \state{Hong Kong SAR}
  \country{China}
}

\author{Jieming Zhu}
\email{jiemingzhu@ieee.org}
\affiliation{%
  \institution{Huawei Noah's Ark Lab}
  \state{Shenzhen}
  \country{China}
}

\author{Jiahao Wu}
\email{jiahao.wu@connect.polyu.hk}
\affiliation{%
  \institution{The Hong Kong Polytechnic University}
  \state{Hong Kong SAR}
  \country{China}
}

\author{Tiandeng Wu}
\email{wutiandeng1@huawei.com}
\affiliation{%
  \institution{Huawei Technolologies Co., Ltd}
  \state{Jiangsu}
  \country{China}
}

\author{Zhenhua Dong}
\email{dongzhenhua@huawei.com}
\affiliation{%
 \institution{Huawei Noah's Ark Lab}
 \state{Shenzhen}
 \country{China}
}

\author{Xiao-Ming Wu}
\authornote{Xiao-Ming Wu is the corresponding author.}
\email{xiao-ming.wu@polyu.edu.hk}
\affiliation{%
  \institution{The Hong Kong Polytechnic University}
  \state{Hong Kong SAR}
  \country{China}
}

\renewcommand{\shortauthors}{Liu, et al.}

\input{Abstract.tex}




\maketitle

\input{Introduction.tex}

\input{RelatedWork.tex}

\input{Preliminary.tex}

\input{Method.tex}

\input{Experiments.tex}

\input{Conclusion.tex}

\begin{acks}
The authors would like to thank Huiling Shen, Jinhua Song, and Li Jiang from Huawei Shanghai Research Center for their valuable discussions and the anonymous reviewers for their helpful comments. We gratefully acknowledge the support of MindSpore\footnote{\url{https://www.mindspore.cn}}, which is a new deep learning framework used for this research.
\end{acks}

\bibliographystyle{ACM-Reference-Format}
\bibliography{FANS}
\appendix
\input{Appendix}

\end{document}

%% file: Abstract.tex
\begin{abstract}

User-curated item lists, such as video-based playlists on Youtube and book-based lists on Goodreads, have become prevalent for content sharing on online platforms. Item list continuation is proposed to model the overall trend of a list and predict subsequent items. Recently, Transformer-based models have shown promise in comprehending contextual information and capturing item relationships in a list. However, deploying them in real-time industrial applications is challenging, mainly because the autoregressive generation mechanism used in them is time-consuming. In this paper, we propose a novel fast non-autoregressive sequence generation model, namely FANS, to enhance inference efficiency and quality for item list continuation. First, we use a non-autoregressive generation mechanism to decode next $K$ items simultaneously instead of one by one in existing models. Then, we design a two-stage classifier to replace the vanilla classifier used in current transformer-based models to further reduce the decoding time. Moreover, to improve the quality of non-autoregressive generation, we employ a curriculum learning strategy to optimize training. Experimental results on four real-world item list continuation datasets including Zhihu, Spotify, AotM, and Goodreads show that our FANS model can significantly improve inference efficiency (up to 8.7x) while achieving competitive or better generation quality for item list continuation compared with the state-of-the-art autoregressive models. We also validate the efficiency of FANS in an industrial setting. Our source code and data will be available at MindSpore/models\footnote{\url{https://gitee.com/mindspore/models/tree/master/research/recommend}} and Github\footnote{\url{https://github.com/Jyonn/FANS}}.

\end{abstract}

%% file: Introduction.tex
\section{Introduction}

\textbf{Background and  Applications of Item List Continuation.} In many online content platforms, user-curated lists have replaced official lists as an integral part of systematic content sharing. Items among these lists may be organized based on user preference, genre, creator, etc. For example, Zhihu\footnote{Online question-answering service: https://www.zhihu.com} users group multiple answers from machine learning tutorials into a topic (i.e., list), and Spotify\footnote{Online music streaming service: https://www.spotify.com} users organize Rihanna's famous songs into a playlist. 
To assist \textit{list builders} in curating long lists for recommending potential follow-up lists, or to continue to provide relevant item lists after \textit{list consumers} have finished the current list, item list continuation system has come into being, disengaging them from cumbersomely searching for relevant content in a plethora of items.

\input{figures/paradigm}

\textbf{Challenges.} Given an input item list, the list continuation system aims to comprehend the sequential information in the list and generate a list of items that continues the previous one. The key challenges for this task are two-fold: 1) how to predict items based on long-term trends in the sequence rather than a specific previous item and 2) how to return multiple items in one request to reduce server overhead in industrial systems while taking into account the interrelationship and order of items.

\textbf{Previous Works.} 
While modeling item sequences has been widely studied for sequential recommendation~\cite{caser,gru4rec,bert4rec}, there is a gap between sequential recommendation (i.e., next-item prediction on user behavior) and list continuation (i.e., list generation with maintaining the consistency of the user-curated list). Most of existing sequential recommendation models are used as \textit{recallers} for next item recommendation and return top-$K$ candidate items with maximal probabilities. Hence, directly using these models for item list continuation is inappropriate, because they fail to capture the order of items and the relationships between them in a generated list.
To remedy this problem, a straightforward solution is to employ an autoregressive strategy, as illustrated on the left of ~\autoref{fig:paradigm}, to generate the item list in a one-by-one manner. Similar to the language modeling tasks~\cite{bert} in natural language processing (NLP), it takes the item of maximum probability in global classification as the prediction. Existing list continuation methods~\cite{car,hyper} are also designed based on this strategy, which unfortunately cannot be easily deployed in production with high requirements of efficiency. Chances are it may incur a large increase in model inference time that scales linearly with the length of the generated sequence and quadratically with the item vocabulary size. Since the item vocabulary of recommender systems is much larger than the token vocabulary in NLP due to item indivisibility, the softmax classifier will suffer from excessive computational complexity.

\textbf{Present Work.} In this paper, we adopt a non-autoregressive approach to accelerate inference efficiency and improve inference quality for item list continuation. As depicted on the right of ~\autoref{fig:paradigm}, we address the shortcomings of previous models by using a non-autoregressive generation strategy to decode the next $K$ items simultaneously and a hierarchical search method for fuzzy matching followed by precise localization. 

Specifically, we propose a \textbf{FA}st \textbf{N}on-autoregressive \textbf{S}equence generation model, namely \textbf{FANS}, with bidirectional Transformer \cite{attention} as the backbone network. Concretely, it makes the following improvements on the typical autoregressive BERT4Rec \cite{bert4rec} model. 1)
Unlike autoregressive models which only append one $\mask$ token at one time and iterate the process for $K$ times, FANS splices $K$ $\mask$ tokens altogether and goes through the process only once. 
2) We leverage item category information (can be obtained by a simple categorizer if not available) and design a two-stage classifier to hierarchically decode latent representations. Simply put, the two-stage classifier will first predict item category and then predict item over category-specific item vocabulary.
3) To improve the inference quality of non-autoregressive generation, we propose to train our model with curriculum learning~\cite{cl} by using easy samples and hard samples in a progressive manner.
The sample difficulty is defined based on the length of the item sequence to be predicted, as will be introduced in \autoref{sec:method}.

To summarize, our contributions are listed as follows:

\begin{itemize}
    \item We make the first attempt to adopt a non-autoregressive generation approach for item list continuation, which helps to speed up inference efficiency and meet the high requirements of industrial systems. 
    \item We propose a novel model FANS by using non-autoregressive generation and a two-stage classifier for hierarchical search to improve inference efficiency, and a curriculum learning based training scheduler to improve inference quality.
    \item We comprehensively evaluate our FANS model on four real datasets \cite{car} including Zhihu, Spotify, AotM, and Goodreads and through online A/B testing in an industrial system. FANS achieves state-of-the-art performance in most metrics and brings an increase in efficiency up to 8.7x.
\end{itemize}

%% file: figures/paradigm.tex
\begin{figure}[t]
\centering
\includegraphics[width=0.95\linewidth]{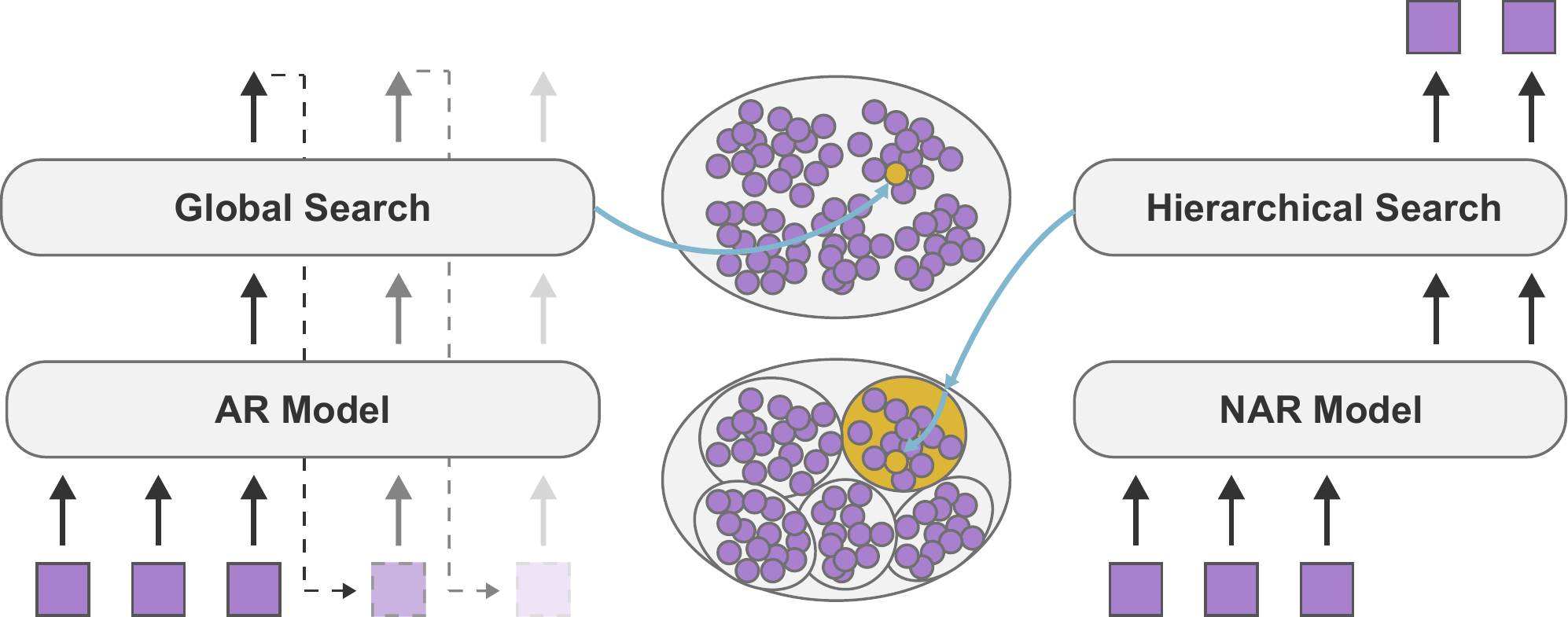}
\caption{Differences between previous autoregressive (AR) models with global search and our proposed non-autoregressive (NAR) model with hierarchical search.}
\Description{Differences between previous autoregressive (AR) models with global search and our proposed non-autoregressive (NAR) model with hierarchical search.}
\label{fig:paradigm}
\end{figure}

%% file: RelatedWork.tex
\section{Related Works \label{sec:rw}}


\textbf{Item list continuation} aims to continue the user-curated list by item sequence generation while maintaining its consistency and coherence, which has received limited study. One line of research focuses on automatic playlist continuation~\cite{chen2018recsys}, which has gained much attention~\cite{volkovs2018two,gatzioura2019hybrid,tran2019adversarial}. However, these methods are often tailored for music playlists.

For general item list continuation, the pioneer works~\cite{mcfee2011natural,mcfee2012hypergraph,chen2012playlist} model item list via Markov chain. Recently, CAR~\cite{car} is proposed to use a consistency-aware gating network to explicitly model item consistency features based on SASRec~\cite{sasrec}. HyperTeNet~\cite{hyper} uses hyper-graph and Transformer network for personalized list continuation by constructing a $K$-NN graph for user, item, and list respectively. Nevertheless, these methods generate lists in an autoregressive manner, which hinders their deployment in industrial systems due to slow inference efficiency.

\textbf{Sequential recommendation} has been increasingly popular in recent years~\cite{rabiner1986introduction,caser,gru4rec,zhou2019online}. Examples include CNN-based Caser \cite{caser}, RNN-based GRU4Rec~\cite{gru4rec}, and recent Transformer-based methods~\cite{he2019hierarchical,sasrec,bert4rec,side1,car,liu2022prec,fan2022sequential,chen2022intent}.
These sequential recommendation models are used as recallers for next-item prediction and return top-$K$ candidate items with maximal probabilities. Thus, they cannot be directly used for list continuation as they don't capture the item order and relationships within a generated list.

One straightforward solution to bridge the gap between sequential recommendation and item list continuation is to leverage autoregressive generation, not item recalling, on these recommendation models for list generation. However, it will also bring the slow inference efficiency problem due to the autoregressive nature.

In this paper, our FANS model adopts non-autoregressive generation and uses a two-stage classifier to accelerate inference efficiency.


\textbf{Non-autoregressive generation} was initially proposed in the area of neural machine translation~\cite{maskpredict,wang2019non,guo2020fine,liu2020task} to speed up inference efficiency and has attracted widespread attention in neural language processing. Mask-Predict~\cite{maskpredict} employs a conditional masked language modeling task to iteratively refine the generated sequence. \citeauthor{wang2019non}~\cite{wang2019non} propose to leverage auxiliary regularization to improve inference quality. There are some recent works~\cite{guo2020fine} utilizing curriculum learning for non-autoregressive generation. Curriculum learning mimics comprehending sequences in human curriculum, training machine learning models starting with easier samples and gradually adding harder samples. It helps improve the generalization ability and convergence speed of the model~\cite{wang2021survey}.

Though non-autoregressive generation and curriculum learning have been explored in the domain of natural language processing, we make the first attempt to use them for item list continuation.
Besides, as the item vocabulary size of list continuation (up to 240K in industrial systems) is much larger than that of natural language processing (approximately 30K), we further design a two-stage classifier to address the inefficiency of ordinary classifier caused by the large vocabulary size.

%% file: Preliminary.tex
\section{Preliminaries \label{sec:preliminary}}

\textbf{Notations and Problem Statement.} Before delving into the details of our proposed FANS model, we first introduce basic notations and formally define the item list continuation problem. Let $\mathcal{V}=\left \{ v_1, v_2, \cdots, v_M \right \} $ denote the item vocabulary where $M$ is the number of items, and each item $v_i$ belongs to a category 
$c_{v_i}$
where $c_{v_i} \in \mathcal{C} = \left \{ 1, 2, \ldots, N \right \}$ is the category id and $N$ is the total number of categories. 
We use $\mathcal{W}^j$ to represent the item set of category $j$ and $M^j$ to denote the size of $\mathcal{W}^j$. Note that $M = \sum^{N}_{j=1}M^j$.
We use $\left ( \front, \rear \right ) \in \mathcal{D}$ to represent an input-target pair in dataset $\mathcal{D}$, where $\front = [ x_1, x_2, \cdots, x_{|\front|}]$ ($x_i\in \mathcal{V}$) is the input list built by an anonymous user and $\rear = [ y_1, y_2, \cdots, y_{|\rear|}]$ ($y_i\in \mathcal{V}$) is the target list, with $|\front|$ and $|\rear|$ being the length of the input and target lists respectively. 
Formally, item list continuation can be defined as: given the input list $\front$, predict the target list $\rear$ that is a subsequent of $\front$, which can be formulated as maximizing the probability
\begin{equation}
    p\left(\rear^\prime = \rear|\front \right),
\end{equation}
where $\rear^\prime$ represents any possible list of length $|\rear|$.

\textbf{Item Categorizer.} For datasets without categorical information, we can first classify the items using a \textit{categorizer}. The categorizer can be of various kinds. In this paper, we take a simple yet effective approach. We first use word2vec~\cite{word2vec} to model item sequences and obtain item representations. Then, we apply $K$-means clustering~\cite{hartigan1979algorithm} to divide the items into $N$ categories.

%% file: Method.tex
\section{Proposed Method\label{sec:method}}

\input{figures/overview}


\subsection{Overview}

Item sequence modeling has been extensively studied for sequential recommendation, and considerable success has been achieved with recent Transformer-based models such as SASRec~\cite{sasrec}, CAR~\cite{car}, and BERT4Rec~\cite{bert4rec}. While SASRec and BERT4Rec have been successfully used in sequential recommender systems, they are treated as \textit{recallers} to obtain the top-$K$ candidate items with maximal probabilities, through the modeling of next-one item prediction. To tackle the item list continuation task that requires capturing the order and relationships of items in the target list, a straightforward way is to take these models as \textit{generators} and perform sequence generation in an autoregressive manner, which is to predict
\begin{equation}
    p\left(\rear^\prime = \rear|\front \right) = \prod^{|\rear|}_{i=1} p\left(y_i^{\prime} = y_i|\front, \rear_{1:i-1} \right).
\end{equation}

However, these autoregressive methods need to decode items one by one (via next-one item prediction by global classifier), which hinders their deployment in industrial scenarios due to low inference efficiency. To meet the high requirements of efficiency in industrial systems, we introduce a non-autoregressive method, called FANS, for parallel sequence generation, which is to predict
\begin{equation}
    p\left(\rear^\prime = \rear|\front \right) = \prod^{|\rear|}_{i=1} p\left(y_i^\prime = y_i|\front \right).
\end{equation}

\autoref{fig:overview} illustrates the overall architecture of our FANS model, including four basic components, i.e., sample masker, embedding layer, Transformer encoder, and two-stage classifier. 
Particularly, FANS takes the input-target list pair with category information as input. The sample masker first operates on the input-target pair, masks part of the items, and generates token-level samples at different difficulty levels. 
Next, the embedding layer projects token sequences with sparse vectors to dense embedding sequences. 
Then, the Transformer encoder models the embedding sequence to learn contextual item representations. 
Finally, the two-stage classifier hierarchically decodes the output embedding and calculates the sample loss.

Importantly, to improve the inference quality of non-autoregres\-sive generation, we employ a curriculum learning based training strategy to optimize training. In effect, it is a training scheduler, controlling sample difficulty levels in sample masker.

\subsection{Non-autoregressive Generation}

To decode the target list all at once, we employ the non-auto\-regressive generation mechanism, which has been explored and used~\cite{guo2020fine} for neural machine translation. More precisely, we use Transformer to model the consecutive sub-sequences, i.e., the input and target lists, where the target list can be regarded as labels. Following BERT4Rec~\cite{bert4rec}, we apply the \textit{Cloze}~\cite{taylor1953cloze} task to mask the label sequence and make the prediction of all items at once, based on the contextual information.

To better match this goal, we use a truncated masking scheme on the target list, i.e., only the sub-list at the end of the list will be masked. Meanwhile, to take advantage of the Cloze task used in BERT~\cite{bert} for pre-training, we apply a random masking scheme on the input list to further enhance contextual modeling. Hence, we name our training objective hybrid item masking.

\textbf{Sample Masker.} The sample masker processes the input-target pair according to hybrid item masking, served to generate samples at different difficulty levels to train the FANS model. 

As shown in \autoref{fig:overview}, each sample, i.e., an input-target list pair, contains an item list (in purple) and a category list (in yellow). For the front item list, we randomly mask $\rho_r$ proportion of items and perform one of the following three masking operations.
For each item $v_i$, there is 1) a $\beta_m$ probability of being replaced with a special token $\mask$; 
2) a $\beta_r$ probability of being substituted with a random item $v_r$ from the vocabulary $\mathcal{V}$; 
and 3) a $\beta_u$ probability of remaining unchanged, where $\beta_m + \beta_r + \beta_u = 1$.
For the target item list, we mask the last $\rho_t$ proportion of the items and replace each with a special $\mask$ token. Meanwhile, the input and target category lists will be adjusted by the following rule. If the token $v_i$ is masked during the previous procedure, the corresponding category token $c_i$ will be replaced with a special $\pad$ token.

Additionally, we use $\rho_t$ to define \textit{sample difficulty}: when $\rho_t$ approaches $0$, the sample is easier; when $\rho_t$ is close to $1$, the sample is harder.

\textbf{Embedding Layer.} The embedding layer first constructs four types of token sequences (i.e., item, category, positional, and segment token sequence), and then projects them from one-hot vectors of length of the vocabulary size into dense low-dimensional vectors of a unified length $d$ (i.e., embedding dimension), and finally integrates the four token embeddings into one input embedding for the Transformer encoder.

The item token sequence is the splicing of the input and target item list. Typically, we add a special $\cls$ token at the beginning of the sequence, and the input and target list is appended with another special $\sep$ token. For clarity, $\cls$ and $\sep$ will be omitted in the subsequent description. 
After concatenating, the item token sequence is represented as:
\begin{equation}
    \bm u = [x_1, \cdots, \mask, \cdots,  x_{|\front|}, y_1, \cdots, \mask, \cdots\mask].
\end{equation}
Similarly, the category token sequence can be constructed as:
\begin{equation}
    \bm c = [c_{x_1}, \cdots, \pad, \cdots,  c_{x_{|\front|}}, c_{y_1}, \cdots, \pad, \cdots\pad].
\end{equation}
Following BERT~\cite{bert}, we also use positional token sequence $\bm p = [0, 1, 2, \ldots, l - 1]$ to carry sequence location information insensitive to the Attention mechanism, and segment token sequence $\bm s = [0, 0, \cdots, 0, 1, \cdots, 1]$ to distinguish input and target lists, where $\bm u$, $\bm c$, $\bm p$ and $\bm s$ have the same length $l$.

Next, we build four consecutive token embedding matrices $\bm X_u \in \mathbb{R}^{(M + 2) \times d}$, $\bm X_c \in \mathbb{R}^{(N + 2) \times d}$, $\bm X_p \in \mathbb{R}^{l \times d}$, and $\bm X_s \in \mathbb{R}^{2 \times d}$. Note that the $\mask$ and $\pad$ tokens will be recognized by $\bm X_u$ and $\bm X_c$, and transformed into a learnable embedding and a fixed zero embedding, respectively.

The final input embedding can be integrated as follows:
\begin{equation}
    \bm E^0 = \bm A_u + \bm A_s + \bm A_p + \bm A_s \in \mathbb{R}^{l \times d}
\label{eq:add}
\end{equation}
where $\bm A_u$, $\bm A_c$, $\bm A_p$ and $\bm A_s$ are item, category, positional, and segment token embeddings after transformation.

\textbf{Transformer Encoder.} Our FANS model uses the Transformer network as the backbone, which is a stack of Transformer layers. Each Transformer layer holds individual weights but is unanimous in structure. Encoded by multiple Transformer layers, the output sequence $\bm E^H$ can be calculated by:
\begin{equation}
    \bm E^h = \mathrm{TRANS}^h \left( \bm E^{h-1}\right), h = 1, 2, \ldots, H,
\end{equation}
where $H$ is the number of Transformer layers, and $\mathrm{TRANS}^h$ is the $h$-th Transformer layer.

\subsection{Two-stage Classifier}

It is well known that the MLM task adopts a multi-layer perceptron classifier (called vanilla classifier) that directly infers the probability distribution over the entire item vocabulary $\mathcal{V}$ by the softmax function, defined as:
\begin{equation}
    \mathrm{C}_{mlm}: \mathbb{R}^d \rightarrow \mathbb{R}^M.
\end{equation}
As the vocabulary grows, the computational complexity and space complexity will increase accordingly, 
having a dominating effect on inference time. For item list continuation, the item vocabulary is much larger than that used in the NLP domain, so directly using the vanilla classifier makes it hard for optimization and is time-intensive. To speed up classification efficiency, some works~\cite{joulin2016bag} utilize Huffman encoding algorithm to construct a frequency-based binary tree for hierarchical search. However, in the scenario of item list continuation, the search criterion should be based on content similarity rather than word frequency. Therefore, we design a two-stage classifier (TSC), to hierarchically decode latent embeddings by performing first-level classification to narrow down the search space of items for second-level classification. 

Precisely, the TSC module consists of one category classifier and one local classifier hub that includes $N$ local classifiers, which can be formulated as:
\begin{equation}
    \mathrm{C}_{cat}: \mathbb{R}^d \rightarrow \mathbb{R}^N,
\end{equation}
\begin{equation}
    \mathrm{C}_{loc}^i: \mathbb{R}^d \rightarrow \mathbb{R}^{M^i}.
\end{equation}

As illustrated in \autoref{fig:overview}, for each latent embedding $\bm E^H_j$ of masked token $u_k$ in sample $\bm t$, it will be passed through the category classifier to predict the item category $c_{u_k}$, and the loss function can be defined as:
\begin{equation}
        L_\mathrm{cat} = -\frac{1}{n} \sum^n_{i=1} \sum_{\bm z \in \bm t} \log \left(\mathrm{C}_{cat}\left(\bm E^H_j\right)_{c_{u_k}}\right),
\end{equation}
where $\bm z$ is the triplet $\left(u_k, c_{u_k}, \bm E^H_j\right)$ (i.e., one of the masked items, its category, and the corresponding latent embedding) in sample $\bm t$ and $n$ is the batch size. If $c_{u_k}$ gets the maximum probability over all categories, the local classifier hub will select the $c_{u_k}$-th local classifier to project the latent embedding to the $c_{u_k}$-th category space $\mathcal{W}^{c_{u_k}}$. The loss function of the $c_{u_k}$-th local classifier can be expressed as:
\begin{equation}
    \Phi\left(\bm z\right) =  1 - \left|\mathrm{sgn} \left(\underset{i \in \left(1, \ldots, N\right)}{{\arg \max} \, \mathrm{C}_{cat}\left(\bm E^H_j\right)_{i}} - c_{u_k} \right)\right|,
\end{equation}
\begin{equation}
    L_\mathrm{loc}^{c_{u_k}} = -\frac{
    \sum\limits^{n}_{i=1} \sum\limits_{\bm z \in \bm t} \Phi\left(\bm z\right) \log \left(\mathrm{C}_{loc}^{c_{u_k}}\left(\bm E^H_j\right)_{u_k}\right)
    }{
    \sum\limits^{n}_{i=1} \sum\limits_{\bm z \in \bm t} \Phi\left(\bm z\right) + \epsilon}.
\end{equation}
where $\Phi\left(\bm z\right)$ is a boolean detector of whether the category classifier yields the correct answer, $\mathrm{sgn}$ is the sign function, and $\epsilon$ is a small real number to prevent the denominator from being $0$. Finally, the loss function can be represented as:
\begin{equation}
    L = L_\mathrm{cat} + \frac{1}{N} \sum^{N}_{i=1} L_\mathrm{loc}^i.
\end{equation}

\input{figures/strategy}

\subsection{Curriculum Learning Based Training \label{sec:ts}} 

While non-autoregressive generation can significantly improve inference efficiency, it may lead to decrease in inference quality. To improve the inference quality of non-autoregressive generation, we design a curriculum learning based training strategy, through controlling sample difficulty levels in sample masker.

\autoref{fig:strategy} illustrates the naive training strategy and our proposed step-wise training strategy, each of which corresponds to a training scheduler. \textbf{Naive scheduler} simply lets the sample masker produce batches of hard samples at all training epochs. Thus, the sample difficulty $\rho_t$ is always 1. \textbf{Step-wise scheduler} opts for an easy-to-hard learning approach, which asks sample masker to generate batches of samples at a certain difficulty level in different training stages. The scheduler first divides all training epochs into units of equal length, and each unit (which consists of a number of training epochs) is considered as \textit{a curriculum step}. Within each unit, the sample difficulty $\rho_t$ is constant. $\rho_t$ increases as training progresses, i.e., harder samples will be used to train the FANS model.

%% file: figures/overview.tex
\begin{figure}[t]
\centering
\includegraphics[width=1.0\linewidth]{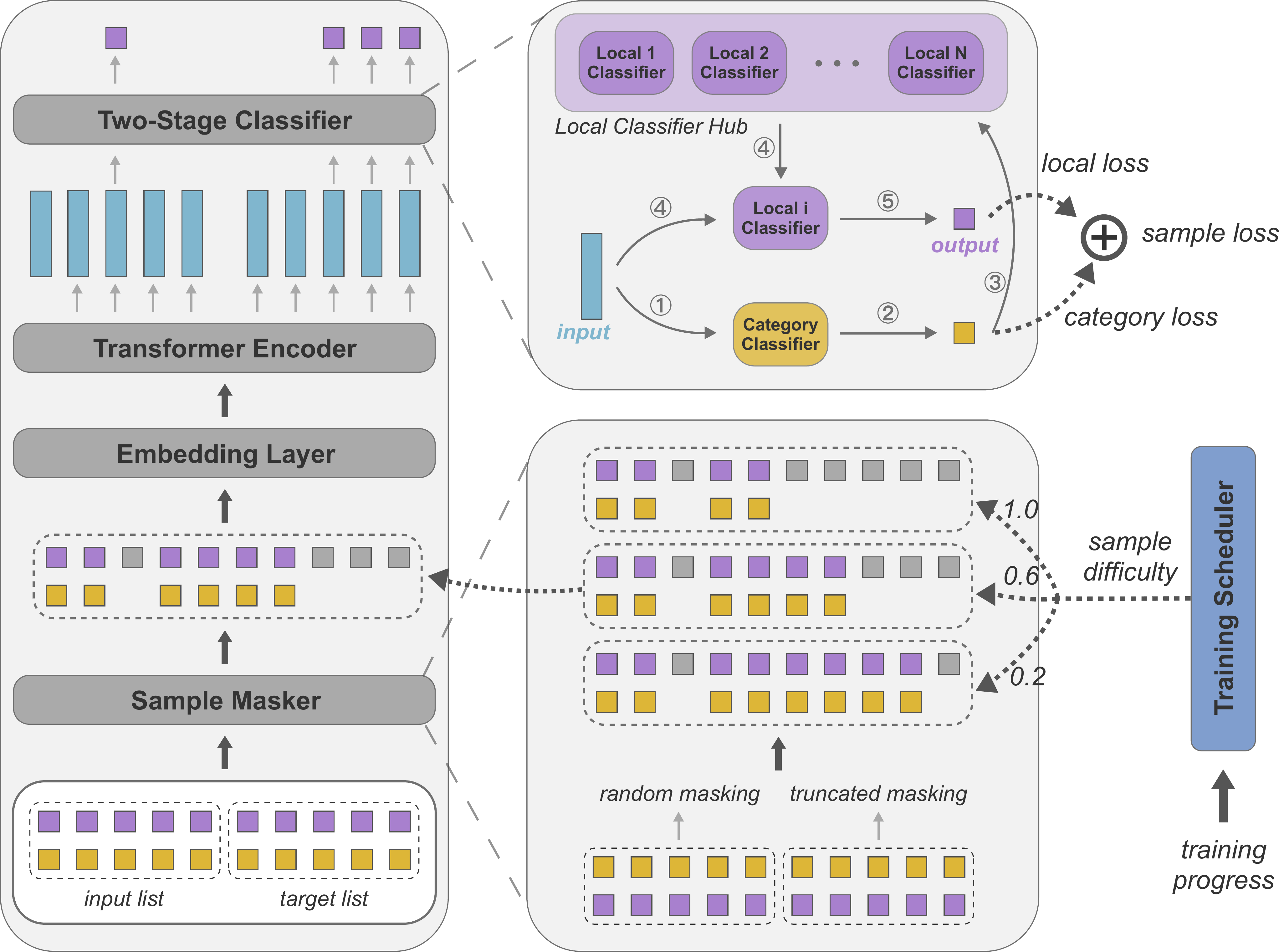}
\caption{Framework of our FANS model. Purple and yellow squares represent item and category tokens respectively, grey squares represent masked tokens, and blue bars indicate latent embeddings. The category tokens of masked items are set to $\pad$ tokens, which are not shown.}
\Description{Framework of our FANS model. Purple and yellow squares represent item and category tokens respectively, grey squares represent masked tokens, and blue bars indicate latent embeddings. During training, a batch of samples will be processed by the sample masker, the embedding layer, and the Transformer encoder, and finally estimated by the two-stage classifier. Meanwhile, the training scheduler will control sample difficulty levels and weights according to training progress.}
\label{fig:overview}
\end{figure}

%% file: figures/strategy.tex
\begin{figure}[t]
\centering
\includegraphics[width=0.9\linewidth]{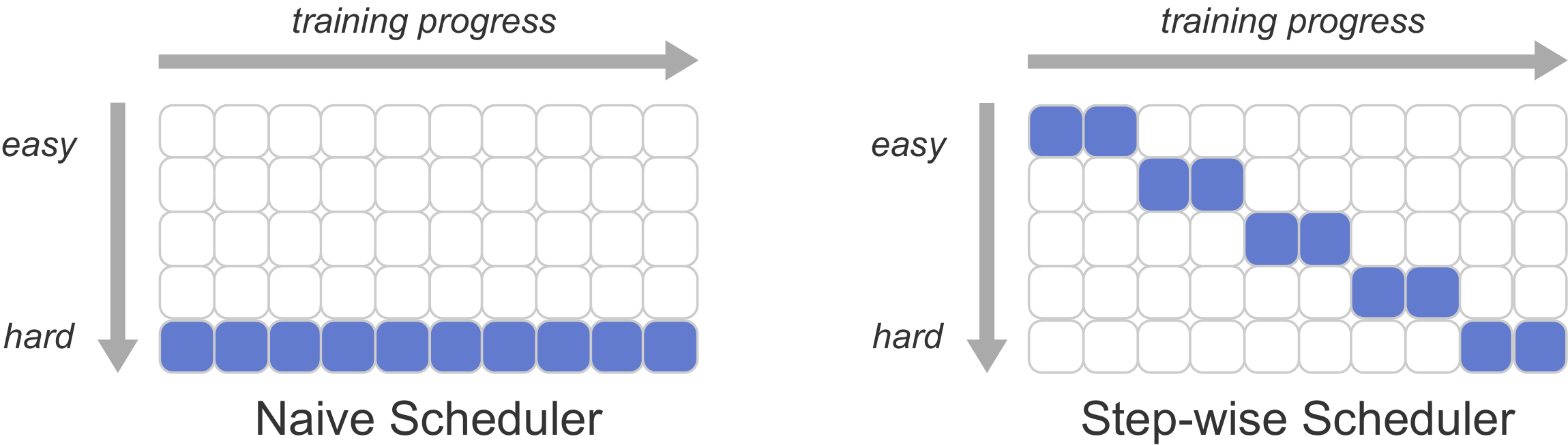}

\caption{Naive scheduler and proposed step-wise scheduler. Each square represents the sample weight at a different difficulty level, which is used for loss calculation during training. Precisely, at each training epoch, naive scheduler converts all training samples to hard samples, while step-wise scheduler converts all training samples to some difficulty level that changes gradually from easy to hard as training goes on.}
\Description{Naive scheduler and proposed step-wise scheduler. Each square represents the sample weight at a different difficulty level, which is used for loss calculation during training. Precisely, at each training epoch, naive scheduler converts all training samples to hard samples, while step-wise scheduler converts all training samples to some difficulty level that changes gradually from easy to hard as training goes on.}
\label{fig:strategy}
\end{figure}

%% file: Experiments.tex
\section{Experiments \label{sec:ex}}


\begin{table}[]

\centering
\renewcommand\arraystretch{1.1}

\caption{\label{tab:statistics}Dataset statistics.}
\resizebox{\linewidth}{!}{
\begin{tabular}{ccccc}
\toprule
 & \textbf{Zhihu} & \textbf{Spotify} & \textbf{AotM} & \textbf{Goodreads} \\ \midrule
\textbf{$\#$List}s & 18,704 & 72,152 & 12,940 & 15,426 \\
\textbf{$\#$Items} & 36,005 & 104,695 & 6,264 & 47,877 \\
\textbf{$\#$Categories} & 10 & 10 & 10 & 10 \\
\textbf{$\#$Interactions} & 927,781 & 6,809,820 & 162,106 & 1,589,480 \\
\textbf{Avg. $\#$ items per list} & 49.59 & 94.38 & 12.53 & 103.04 \\
\textbf{Range $\#$ items per list} & $10\sim200$ & $20\sim300$ & $10\sim60$ & $20\sim300$ \\
\textbf{Density} & 0.138\% & 0.089\% & 0.200\% & 0.215\% \\
\bottomrule 
\end{tabular}
}

\end{table}

\subsection{Experimental Setup}

\textbf{Datasets.} Following~\cite{car}, we conduct offline experiments on four real-world item list continuation datasets, i.e., AotM, Zhihu, Spotify, and Goodreads. The AotM dataset is proposed by~\cite{mcfee2012hypergraph}, and the other three are crawled and built by~\cite{car}. AotM and Spotify are automatic playlist continuation datasets collected from Art of the Mix\footnote{http://www.artofthemix.org/} and Spotify respectively. Zhihu is a question-and-answer community, and Goodreads\footnote{https://www.goodreads.com/} is a website for book recommendation. The statistics of the four datasets are summarized in \autoref{tab:statistics}. Data preprocessing and implementation details are provided in the appendix.

Since the items in the datasets lack categorical features, we cluster them based on sequence information. First, we filter out items with word frequency less than 10 and apply the Continuous Bag-of-Words (CBOW)~\cite{word2vec} model to learn item latent vectors. Specifically, we set the embedding dimension as 64 and the window size as 2. The training objective is to predict the current item based on $2 + 2$ surrounding items. Second, we utilize the $K$-means clustering algorithm~\cite{hartigan1979algorithm} to cluster the item vectors. We apply Euclidean distance to describe the distance between two item embedding vectors and divide items into 10 categories. Third, we iteratively perform the following two operations until the data is fixed: 1) remove items with a frequency less than 10 from all lists; 2) truncate or filter the item list according to the maximum and minimum lengths w.r.t. each dataset. We evenly divide an eligible list into two segments, i.e., the input and target lists. Fourth, the mapping of items and categories is established according to the final lists. The lists are divided into training, validation, and testing sets by a ratio of 8:1:1.

\input{tables/bigtable}

\textbf{Baselines and Variants of Our Method.} To verify the efficiency and effectiveness of our proposed FANS model, we compare with state-of-the-art sequential recommendation methods and item list continuation models. Notice that since we leverage item category information in our FANS model, for a fair comparison, we also inject category embedding into item embedding (by vector addition as in~\autoref{eq:add}) for all compared models. We verify the effectiveness of category embedding in \autoref{fig:use}. 
In the following, we briefly introduce the baselines and variants of our method.

We compare with the following autoregressive models: \textbf{Caser}~\cite{caser} utilizes the convolutional neural network (CNN) to vertically and horizontally model the local sequence. \textbf{GRU4Rec}~\cite{gru4rec} employs the gated recurrent unit (GRU) network on the sequential recommendation. \textbf{SASRec}~\cite{sasrec} is a Transformer-based sequential recommendation model which leverages unidirectional information in list modeling. \textbf{BERT4Rec}~\cite{bert4rec} is a state-of-the-art model which also uses a Transformer as the backbone. In the training phase, it uses a self-supervised MLM task to comprehend contextual information, and the $\mask$ token is appended at the end of the input list for prediction during the inference phase. \textbf{CAR}~\cite{car} is a Transformer-based item list continuation model based on SASRec. It proposes a consistency-aware gating network to explicitly model item consistency features. Since only anonymized user-generated lists are stored in our re-generated datasets, we drop the GUPM (General User Preference Model) network which utilizes user-id features.

We further design a non-autoregressive version of BERT4Rec: \textbf{\nagbert} (which is equivalent to a naive variant of FANS, namely, \textbf{\fansB}). Since there is no non-autoregressive model proposed for item list continuation, we transform the BERT4Rec model into a non-autoregressive model by appending multiple $\mask$ tokens at the end of the list.

We evaluate the following variants of our proposed FANS model: \textbf{\fansV} is accelerated by the non-auto\-regressive mechanism with a step-wise scheduler. Note that \textbf{\fansB} uses the naive scheduler. \textbf{\fansN} is accelerated by the non-auto\-regressive mechanism with the naive scheduler and adopts a two-stage classifier. \textbf{\fansS} is accelerated by the non-auto\-regressive mechanism with a step-wise scheduler and adopts a two-stage classifier. 

\textbf{Evaluation Protocols.} To evaluate the quality of the recommended list, we follow previous work~\cite{shi2019deep} and adopt Normalized Discounted Cumulative Gain~\cite{ndcg} (NDCG@k) and Hit Ratio (HR@k) metrics to measure the performance of each model. In this work, we set $k = \{5, 10\}$. 
To evaluate model efficiency, we measure latency, i.e., inference efficiency per sample in milliseconds, obtained by dividing the total time spent for inference by the total number of samples in the test set. We set the batch size to 1 and average the results of five runs with a NVIDIA GeForce RTX 3090 graphics card.

\input{tables/strategy}

\subsection{Comparison with State-of-the-art Methods}

We compare the performance of our FANS model with state-of-the-art methods on four item list continuation datasets in both inference quality and efficiency.
\autoref{big-table} summarizes the overall performance of all models. The ``Speedup'' metric is computed as the ratio of the latency (ms) of the present model and that of the BERT4Rec model, a state-of-the-art sequential recommender. From the results, we can make the following observations.

First, by comparing the five autoregressive models (in blue), we can conclude that the Transformer-based BERT4Rec model performs relatively well on the four metrics on inference quality (i.e., NDCG@\{5, 10\} and HR@\{5, 10\}).
Specifically, BERT4Rec performs best on 2 metrics and second best on 1 metric on the AotM dataset.

Second, autoregressive models (in blue) are much less efficient and also less effective than models accelerated with only the non-auto\-regressive mechanism (in orange). \fansV is also only accelerated with the non-autoregressive mechanism, so its latency is equivalent to that of \nagbert. However, with the step-wise curriculum learning training strategy, it outperforms \nagbert and achieves state-of-the-art performance in 13 out of 16 cases. Compared with BERT4Rec, \fansV not only achieves significant speedups (from 2.4x to 3.0x) on all four datasets but also excels in most cases. The reasons may be 1) the results generated in autoregressive manner are easily disturbed by previously generated items, causing the sequence to deviate from the normal trend; 2) long-term trend forecast for items generated by non-autoregressive methods is more accurate with the help of curriculum learning based training scheduler. However, when the item vocabulary size $M$ is too large (e.g., 104K items of Spotify dataset), \fansV still cannot meet the real-time requirements (i.e., latency less than 50ms).

Third, our FANS model, accelerated with the non-autoregressive mechanism and two-stage classifier (in green), achieves much higher inference efficiency. 
Compared with \nagbert, FANS leverages a two-stage classifier to perform hierarchical decoding on latent embeddings to further accelerate inference. For example, on the Spotify dataset, FANS can reach 8.7 times speedup, compared to the 2.4 times speedup by \nagbert. Notably, \fansN without curriculum learning performs much worse than \fansS trained with step-wise scheduler. Compared with \fansV, the two-stage classifier achieves a substantial improvement in inference efficiency.

Fourth, on small datasets, \fansV is sufficient for real-time inference and achieves the best inference quality. When dataset size grows (especially vocabulary size grows), the trade-off needs to be considered, and for most of the time, it is acceptable to sacrifice a slight reduction in inference quality for inference efficiency.

\input{plots/use}
\input{plots/recall}
\input{tables/cluster}

\subsection{Ablation Studies}

Here, we study the effect of different components including the training schedulers and the two-stage classifier on the performance of FANS. Experiments of \autoref{ablation-strategy} 
are conducted on the Zhihu dataset. We vary the \textit{curriculum step} parameter by $\{3, 5, 7, 10\}$. Besides, we compare the performance of utilizing the vanilla classifier proposed by BERT~\cite{bert} and our proposed two-stage classifier. Based on the results, we can draw the following observations. \textbf{First}, for the training schedulers, the \textit{curriculum step} of the step-wise scheduler needs to be properly tuned, otherwise the training performance would be worse than that of the naive scheduler. When the curriculum step is set to 5, the training performance reaches the best. \textbf{Second}, by comparing the vanilla classifier with the two-stage classifier, we notice that the vanilla classifier outperforms the two-stage classifier on most metrics. It is probably because the two-stage classifier narrows down the item search space which improves inference efficiency but preventing correct items in other categories from being selected.

Furthermore, \autoref{fig:recall} compares the performance among recall-based BERT4Rec (i.e., BERT4Rec$_\texttt{R}$), autoregressive-based BERT4Rec, and non-autoregressive-based \fansS models. Note that BERT\-4Rec$_\texttt{R}$ and BERT4Rec use the same trained model but with different ways of item generation. BERT4Rec$_\texttt{R}$ is employed as a recaller, performs only one next-item prediction operation, uses the latent embedding to retrieve items, and obtains top-$K$ candidate items with maximal probabilities as a sequence. In contrast, BERT4Rec uses autoregressive generation to obtain the sequence. Although BERT4Rec$_\texttt{R}$ is faster than \fansS, its inference quality is not as good as BERT4Rec, and far worse than \fansS. 

\subsection{Impact of Hyper-parameters on Inference Quality and Inference Efficiency}

First, we study the effect of category information on the performance of FANS.
CAR$^*$ and BERT4Rec$^*$ represent removing categorical features for CAR and BERT4Rec, respectively.
Since the two-stage classifier is based on category information, we use the vanilla classifier as a replacement to construct the model without category feature injection, namely, FANS$^*$. \autoref{fig:use} demonstrates the influence of category features on the inference quality of the three models on four datasets, from which we can make the following observations. First, by comparing CAR$^*$ with CAR, and BERT4Rec$^*$ with BERT4Rec, it can be seen that the inference quality is improved on four datasets, demonstrating the usefulness of category information.
Second, in most cases, FANS$^*$ outperforms the \fansS model, because the two-stage classifier narrows down the item search space which may prevent correct items in other categories from being selected.

Next, we study various hyper-parameter settings that may affect the inference efficiency of FANS.
\autoref{ablation-cluster} demonstrates the influence of item vocabulary size $M$ and category vocabulary size $N$. We can make the following observations. Firstly, note that the inference time of BERT4Rec does not change with the size of category vocabulary. We find that the inference time of our FANS model with one category is always faster than the BERT4Rec model, in which case the two-stage classifier is exactly the same as the vanilla classifier, i.e., FANS$_\texttt{TSC-STEP[N=1]}$ is equivalent to \fansV. 
It suggests that the non-autoregressive model is more efficient than the autoregressive model. Secondly, on each dataset, as the category size grows, the inference time significantly decreases and then gradually increases. 
The reason is that when the category size increases, the time consumption of the first-layer category classifier will increase, while the time consumption of the second-layer local classifier will decrease. 
When the category size reaches a certain threshold, the increase in time consumption of the category classifier will exceed the reduction in time consumption of the local classifier. 
The threshold for the category size differs for each dataset, and the larger the item vocabulary size, the higher the threshold. 
Ideally, we can speed up the Spotify dataset by a factor of 11.9 (when the category vocabulary size $N$ reaches 100) and the Goodreads dataset by 9.2 (when $N$ reaches 20).

\subsection{Assessment of Inference Efficiency in an Industrial System}

We also verify the efficiency of FANS in Huawei's music list continuation system, which serves millions of users daily. The task is to provide follow-up item lists after users have finished the playlist. Due to long item sequences and huge amount of music data (about 240K after filtering), a direct use of 3-layer BERT4Rec model will incur an intolerable delay of more than 300ms (tested with a Tesla P100 PCIe GPU). To improve inference efficiency, we apply our \fansS model. We use the item categorizer to classify music items. We set the number of classes to 10 (i.e., FANS$_\texttt{[N=10]}$) and 100 (i.e., FANS$_\texttt{[N=100]}$) and find the inference efficiency decreased from 300ms to 42ms (7.1x) and 28ms (10.7x), respectively. Compared with FANS$_\texttt{[N=10]}$, the inference quality of FANS$_\texttt{[N=100]}$ is only 1.2\% lower, while the inference speed is 1.5 times faster, which is completely acceptable in the application scenario.

%% file: tables/bigtable.tex
\begin{table*}[]
\centering
\renewcommand\arraystretch{1.1}
 
\caption{Comparison of our FANS model with state-of-the-art methods. Blue circle (\ag{}) indicates an autoregressive model. Orange circle (\nag{}) indicates using non-autoregressive generation. Green circle (\tsc{}) indicates using both non-autoregressive generation and the proposed two-stage classifier. The ``Speedup'' metric is computed as the ratio of the latency (ms) of the present model and that of BERT4Rec. We bold the best result and underline the second best. Results with latency above 50ms are in red (i.e., \fansuac{fail to meet} the efficiency requirement), and those below 50ms are in green (i.e., \fansac{meet} the efficiency requirement).}
\label{big-table}

\setlength\tabcolsep{2.5pt}

\resizebox{0.95\linewidth}{!}{
\begin{tabular}{ccccccccccc}
\toprule
 & 
& {\ag{} Caser}
& {\ag{} GRU4Rec}
& {\ag{} SASRec}
& {\ag{} BERT4Rec}
& {\ag{} CAR}
& \begin{tabular}{c} \nag{} \nagbert \\[-0.3em] / \fansB \end{tabular}
& {\nag{} \fansV}
& {\tsc{} \fansN}
& {\tsc{} \fansS}
 \\ 
\midrule

\multirow{6}{*}{
\rotatebox{90}{\textbf{Zhihu}}
}
 & NDCG@5  
    & 0.0169 & 0.0077 & 0.0125 & 0.0226 & 0.0156 
    & 0.0216 & \first{0.0256}
    & 0.0192 & \second{0.0232} \\
 & NDCG@10 
    & 0.0212 & 0.0132 & 0.0194 & 0.0329 & 0.0228 
    & 0.0325 & \first{0.0389}
    & 0.0284 & \second{0.0337} \\
 & HR@5    
    & 0.1862 & 0.1049 & 0.1707 & 0.2477 & 0.1792 
    & 0.2399 & \first{0.2857}
    & 0.2295 & \second{0.2670} \\
 & HR@10   
    & 0.2960 & 0.2255 & 0.3081 & 0.4275 & 0.3248 
    & 0.4202 & \first{0.4819} 
    & 0.3953 & \second{0.4604} \\
 & Latency (ms)
    & \fansuac{88.82} & \fansuac{92.56} & \fansuac{125.74} & \fansuac{120.38} & \fansuac{118.76} 
    & \fansac{42.33} & \fansac{42.33} 
    & \fansac{\first{18.39}} & \fansac{\first{18.39}} \\
 & Speedup 
    & 1.4x & 1.3x & 1.0x & 1.0x & 1.0x
    & 2.8x & 2.8x 
    & \first{6.5x} & \first{6.5x} \\ 
\midrule
 
\multirow{6}{*}{
\rotatebox{90}{\textbf{Spotify}}
}
 & NDCG@5  
    & 0.0277 & 0.0067 & 0.0081 & 0.0211 & 0.0254 
    & 0.0272 & \second{0.0313}
    & 0.0034 & \first{0.0315} \\
 & NDCG@10 
    & 0.0393 & 0.0101 & 0.0113 & 0.0304 & 0.0366 
    & 0.0402 & \first{0.0461}
    & 0.0047 & \second{0.0438} \\
 & HR@5    
    & 0.3477 & 0.1293 & 0.1536 & 0.2811 & 0.3307 
    & 0.3867 & \first{0.4071}
    & 0.0711 & \second{0.3992} \\
 & HR@10   
    & 0.4857 & 0.2110 & 0.2286 & 0.4227 & 0.4755 
    & 0.5502 & \first{0.5729}
    & 0.1188 & \second{0.5552} \\
 & Latency (ms)
    & \fansuac{230.51} & \fansuac{259.18} & \fansuac{240.48} & \fansuac{325.17} & \fansuac{279.07}
    & \fansuac{132.88} & \fansuac{132.88} 
    & \fansac{\first{37.46}} & \fansac{\first{37.46}} \\
 & Speedup 
    & 1.4x & 1.3x & 1.4x & 1.0x & 1.2x 
    & 2.4x & 2.4x 
    & \first{8.7x} & \first{8.7x} \\
 
\midrule

\multirow{6}{*}{
\rotatebox{90}{\textbf{AotM}}
}
 & NDCG@5  
    & 0.0112 & 0.0053 & 0.0104 & 0.0218 & 0.0072 
    & 0.0171 & 0.0214
    & \first{0.0229} & \second{0.0220} \\
 & NDCG@10 
    & 0.0312 & 0.0127 & 0.0662 & \first{0.1008} & 0.0314 
    & 0.0890 & 0.0992
    & 0.0744 & \second{0.0995} \\
 & HR@5    
    & 0.0635 & 0.0317 & 0.0565 & \second{0.1115} & 0.0433 
    & 0.0983 & \first{0.1130}
    & 0.1084 & 0.1053 \\
 & HR@10   
    & 0.2923 & 0.1538 & 0.4308 & \first{0.5538} & 0.2462 
    & 0.5385 & \first{0.5538}
    & 0.4769 & 0.4923 \\
 & Latency (ms)
    & \fansac{10.84} & \fansac{10.84} & \fansac{17.03} & \fansac{18.58} & \fansac{17.80}
    & \fansac{7.70} & \fansac{7.70}
    & \fansac{\first{7.01}} & \fansac{\first{7.01}} \\
 & Speedup 
    & 1.7x & 1.7x & 1.1x & 1.0x & 1.0x 
    & 2.4x & 2.4x
    & \first{2.7x} & \first{2.7x} \\
    
\midrule

\multirow{6}{*}{
\rotatebox{90}{\textbf{Goodreads}}
}
 & NDCG@5  
    & 0.0205 & 0.0074 & 0.0187 & 0.0134 & 0.0232 
    & 0.0257 & \first{0.0334}
    & 0.0276 & \second{0.0293} \\
 & NDCG@10 
    & 0.0278 & 0.0104 & 0.0271 & 0.0204 & 0.0332 
    & 0.0355 & \first{0.0467}
    & 0.0390 & \second{0.0418} \\
 & HR@5    
    & 0.2406 & 0.1154 & 0.2464 & 0.1958 & 0.2789 
    & 0.3104 & \first{0.3891}
    & 0.3171 & \second{0.3268} \\
 & HR@10   
    & 0.3294 & 0.1783 & 0.3632 & 0.3379 & 0.4196 
    & 0.4492 & \first{0.5149}
    & 0.4202 & \second{0.4514} \\
 & Latency (ms)
    & \fansuac{134.38} & \fansuac{165.70} & \fansuac{188.07} & \fansuac{194.55} & \fansuac{233.46}
    & \fansuac{64.92} & \fansuac{64.92}
    & \fansac{\first{24.73}} & \fansac{\first{24.73}} \\
 & Speedup 
    & 1.4x & 1.2x & 1.0x & 1.0x & 0.8x 
    & 3.0x & 3.0x 
    & \first{7.9x} & \first{7.9x} \\
 
\bottomrule
\end{tabular}
}
\end{table*}

%% file: tables/strategy.tex
\begin{table}[t]
\centering
\renewcommand\arraystretch{1.1}

\caption{Influence of the training schedulers on inference quality. The experiments are conducted on the Zhihu dataset. ``naive'' means using the naive scheduler. 
``step$=s$'' means using the step-wise scheduler with $s$ curriculum steps.
}
\label{ablation-strategy}

\setlength\tabcolsep{2.5pt}

\resizebox{0.95\linewidth}{!}{
\begin{tabular}{ccccccc}
\toprule
 \textbf{classifier} & \textbf{scheduler}
 & \begin{tabular}{c} \textbf{naive} \\[-0.3em] / \textbf{step=1} \end{tabular}
 & \textbf{step=3}
 & \textbf{step=5}
 & \textbf{step=7}
 & \textbf{step=10} \\ 
\midrule
\multirow{4}{*}{
\begin{tabular}{c} \textbf{Vanilla} \\[-0.3em] \textbf{Classifier} \end{tabular}
}
 & NDCG@5  & 0.0216 & 0.0232 & 0.0256 & 0.0268 & \first{0.0274} \\
 & NDCG@10 & 0.0325 & 0.0344 & 0.0389 & \first{0.0390} & 0.0370 \\ 
 & HR@5    & 0.2399 & 0.2734 & \first{0.2857} & 0.2840 & 0.2820 \\
 & HR@10   & 0.4202 & 0.4671 & \first{0.4819} & 0.4792 & 0.4577 \\
\hline
\multirow{4}{*}{
\begin{tabular}{c} \textbf{Two-stage} \\[-0.3em] \textbf{Classifier} \end{tabular}
}
 & NDCG@5  & 0.0210 & 0.0226 & \first{0.0232} & 0.0224 & 0.0210 \\
 & NDCG@10 & 0.0312 & 0.0329 & \first{0.0337} & 0.0332 & 0.0310 \\ 
 & HR@5    & 0.2424 & 0.2627 & \first{0.2670} & 0.2611 & 0.2504 \\
 & HR@10   & 0.4101 & 0.4362 & \first{0.4604} & 0.4416 & 0.4201 \\
\bottomrule
\end{tabular}
}
\end{table}

%% file: plots/use.tex
\begin{figure*}
    \centering
    \setlength\tabcolsep{2pt}

    \resizebox{0.9\linewidth}{!}{
    \begin{tabular}{m{0.25\textwidth}m{0.25\textwidth}m{0.25\textwidth}m{0.25\textwidth}}
    \multicolumn{4}{c}{
        \resizebox{1.0\linewidth}{!}{
            \includegraphics{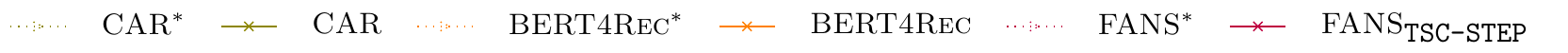}
        }
    } \\
    \begin{subfigure}{0.25\textwidth}
        \resizebox{1.0\linewidth}{!}{
            \includegraphics{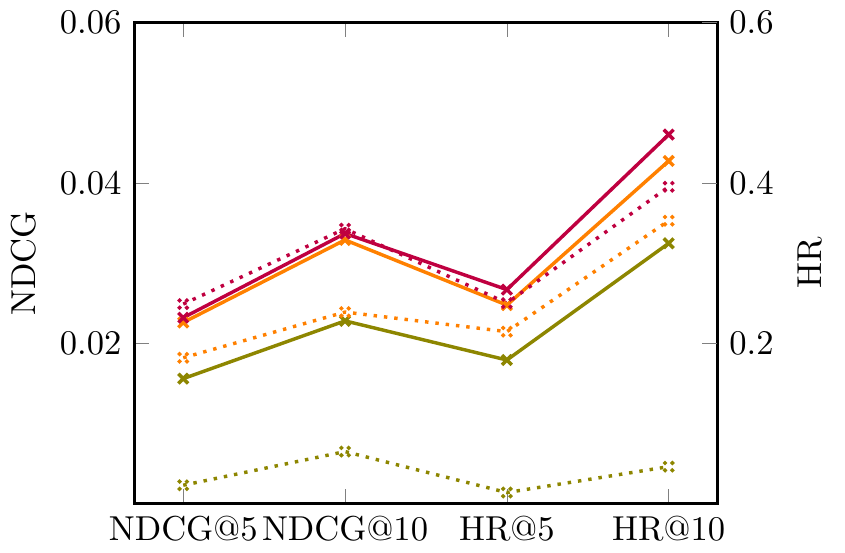}
        }
        \caption{\label{fig:use-zhihu}Zhihu}
    \end{subfigure} & 
    \begin{subfigure}{0.25\textwidth}
        \resizebox{1.0\linewidth}{!}{
            \includegraphics{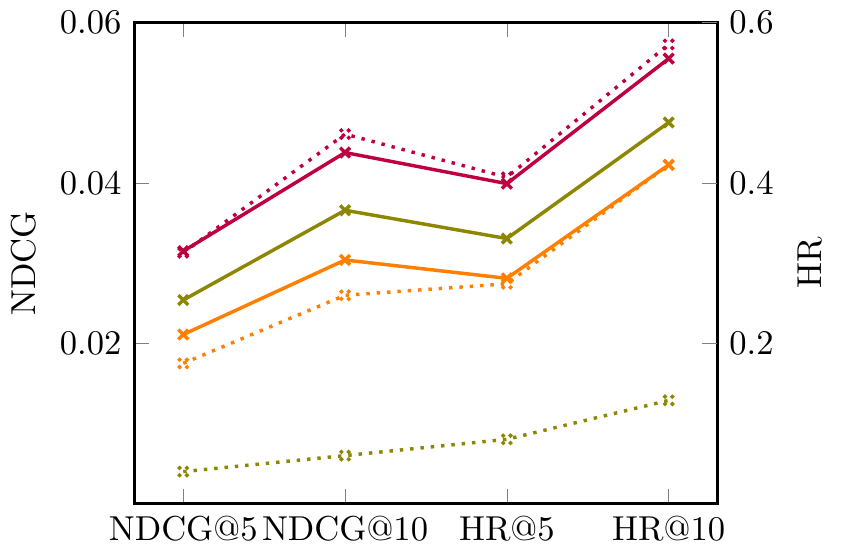}
        }
        \caption{\label{fig:use-spotify}Spotify}
    \end{subfigure} &
    \begin{subfigure}{0.25\textwidth}
        \resizebox{1.0\linewidth}{!}{
            \includegraphics{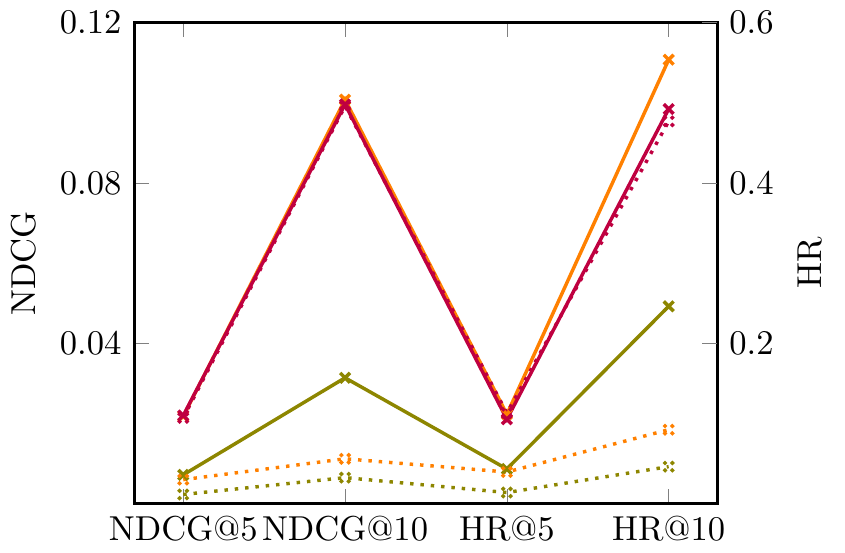}
        }
        \caption{\label{fig:use-aotm}AotM}
    \end{subfigure} & 
    \begin{subfigure}{0.25\textwidth}
        \resizebox{1.0\linewidth}{!}{
            \includegraphics{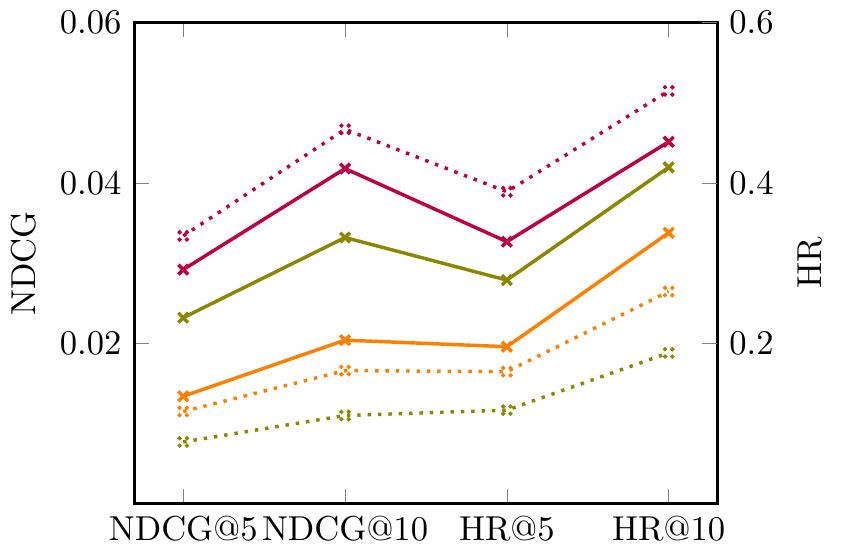}
        }
        \caption{\label{fig:use-goodreads}Goodreads}
    \end{subfigure} 
    \end{tabular}
    }
    
    \caption{\label{fig:use} Influence of the category feature. The solid lines represent models trained with both item list and category list, while dotted lines represent models trained with item list only.}
    \Description{Influence of the category feature. The solid lines represent models trained with both item list and category list, while dotted lines represent models trained with item list only.}
\end{figure*}

%% file: plots/recall.tex
\begin{figure}
    \centering
    \setlength\tabcolsep{0pt}

    \resizebox{0.9\linewidth}{!}{
    \begin{tabular}{m{0.25\textwidth}m{0.25\textwidth}}
    \multicolumn{2}{c}{
        \resizebox{1.0\linewidth}{!}{
            \includegraphics{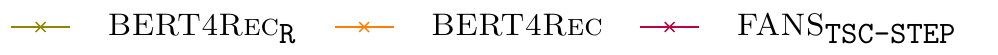}
        }
    } \\
    \begin{subfigure}{0.25\textwidth}
        \resizebox{1.0\linewidth}{!}{
            \includegraphics{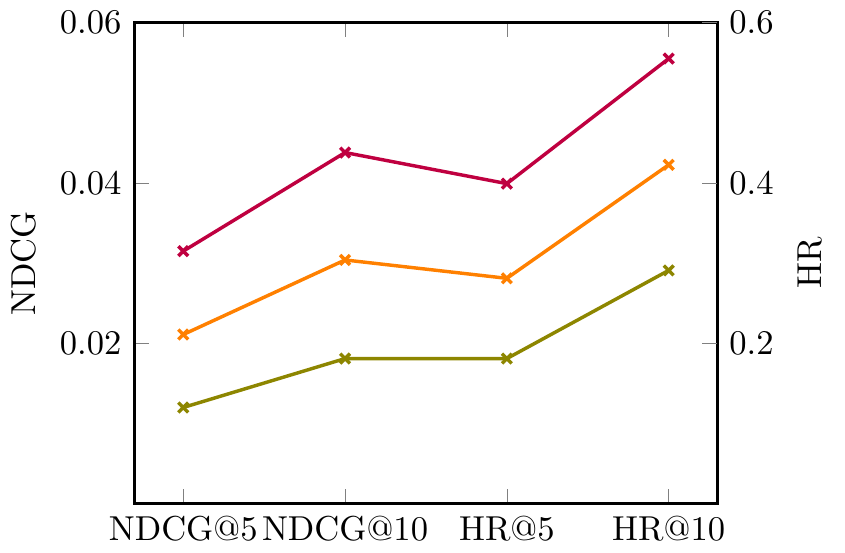}
        }
        \caption{\label{fig:recall-spotify}Spotify}
    \end{subfigure} & 
    \begin{subfigure}{0.25\textwidth}
        \resizebox{1.0\linewidth}{!}{
            \includegraphics{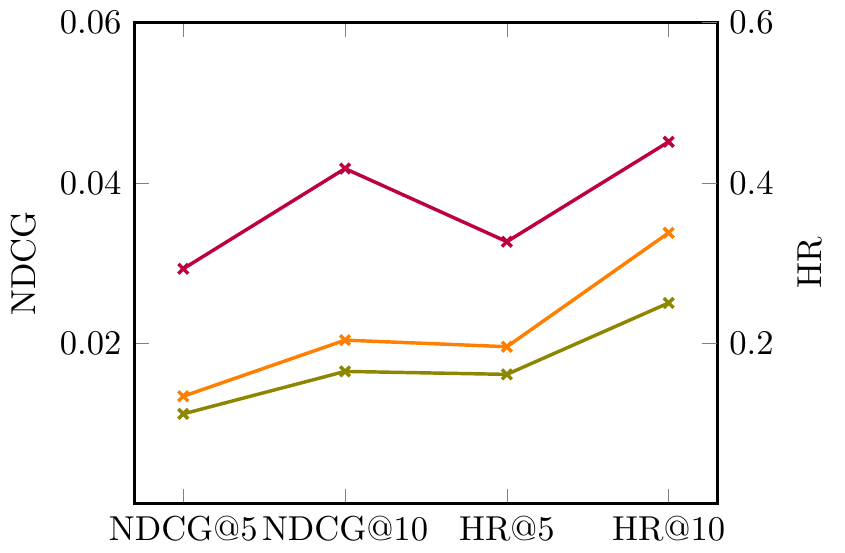}
        }
        \caption{\label{fig:recall-goodreads}Goodreads}
    \end{subfigure}
    \end{tabular}
    }
    \caption{\label{fig:recall} Performance comparison between sequence generation methods (i.e., \fansS, BERT4Rec) and recall-based method (i.e., BERT4Rec$_\texttt{R}$).  
    Note that BERT4Rec and BERT4Rec$_\texttt{R}$ share the same network parameters but generate results in different ways: 
    BERT4Rec continues the list autoregressively, while BERT4Rec$_\texttt{R}$ is used as a recaller to obtain the top-$K$ items with highest prediction probabilities.}
    \Description{Performance comparison between sequence generation methods (i.e., \fansS, BERT4Rec) and recall-based method (i.e., BERT4Rec$_\texttt{R}$).  
    Note that BERT4Rec and BERT4Rec$_\texttt{R}$ share the same network parameters but generate results in different ways: 
    BERT4Rec continue the list autoregressively, while BERT4Rec$_\texttt{R}$ is used as a recaller to obtain the top-$K$ items with highest prediction probabilities.}
\end{figure}

%% file: tables/cluster.tex
\begin{table*}[]
\centering
\renewcommand\arraystretch{1.1}

\caption{Influence of item vocabulary size (M) and category vocabulary size (N) on inference efficiency.}
\label{ablation-cluster}

\setlength\tabcolsep{3pt}

\resizebox{0.9\linewidth}{!}{
\begin{tabular}{ccccccccc}
\toprule
 & \textbf{BERT4Rec} & \multicolumn{7}{c}{\textbf{FANS$_\texttt{TSC}$}} \\
\cmidrule(lr){3-9}
 $N$ & - & 1 (vanilla) & 5 & 10 & 20 & 50 & 100 & 250 \\
\midrule
\textbf{Zhihu ($M=36,005$)}
 & 120.38 (1.0x) & 41.30 (2.9x) & 18.96 (6.3x) & 18.81 (6.4x) & \textbf{17.73 (6.8x)} & 18.46 (6.5x) & 20.50 (5.9x) & 30.82 (3.9x) \\
\textbf{Spotify ($M=104,695$)}
 & 325.17 (1.0x) & 132.14 (2.5x) & 46.22 (7.0x) & 37.46 (8.7x) & 30.90 (10.5x) & 28.14 (11.6x) & \textbf{27.33 (11.9x)} & 36.09 (9.0x) \\
\textbf{AotM ($M=6,264$)}
 & 18.58 (1.0x) & 7.83 (2.4x) & 7.16 (2.6x) & \textbf{7.01 (2.7x)} & 8.22 (2.3x) & 9.80 (1.9x) & 12.39 (1.5x) & 22.51 (0.8x) \\
\textbf{Goodreads ($M=47,877$)}
 & 194.55 (1.0x) & 64.11 (3.0x) & 29.51 (6.6x) & 24.73 (7.9x) & \textbf{21.18 (9.2x)} & 23.22 (8.4x) & 23.42 (8.1x) & 40.50 (4.8x) \\
\bottomrule
\end{tabular}
}
\end{table*}

%% file: Conclusion.tex
\section{Conclusion \label{sec:conclusion}}

To deploy item list continuation models to industrial scenarios with real-time inference requirements, we propose a FANS model to accelerate inference efficiency through non-autoregressive generation and a two-stage classifier strategy for hierarchical decoding. To the best of our knowledge, this is the first attempt to introduce non-autoregressive generation for item list continuation. Furthermore, we leverage a curriculum learning-based training strategy to train our FANS model to improve the quality of non-autoregressive generation. The experimental results on real-world datasets and efficiency evaluation in an industrial system verify the superiority of our FANS model, which can greatly improve inference efficiency while achieving competitive or better generation quality compared with the state-of-the-art methods.


%% file: Appendix.tex
\section{Implementation Details} 



\subsection{Hyper-parameter Selection}

We adopt the Adam optimizer as the gradient descent algorithm. 
The learning rates is set as 0.01. 
The batch size on four datasets is set as 256, and the embedding dimension is set as 64. 
Following BERT~\cite{bert}, we set the random masking ratio $\rho_r$ to 0.15, and set the probabilities of the three masking operations (i.e., $\beta_m$, $\beta_r$, $\beta_u$) to 0.80, 0.10, and 0.10, respectively.
For a fair comparison, we set the number of Transformer layers of Transformer-based models and the number of hidden layers of the GRU4Rec model to both 3.
For the Caser model, we follow the original implementation and settings, and set the max sequence length to 5. 
We set the number of attention heads to 8 for all Transformer-based methods on four datasets.
We consider the \textit{curriculum step} of the step-wise scheduler from \{3, 5, 7, 10\}. Unless otherwise stated, we set the number of curriculum steps as 5.

\subsection{Inference Procedure}

To infer the target list, we perform the following operations:
\begin{enumerate}
    \item Model execution. For non-autoregressive-based methods, the $\mask$ tokens with a length equal to $|\mathbf{y}|$ are added to the input list $\mathbf{x}$, and the trained model produces a $|\mathbf{y}| \times M$ probability matrix all at once. On the other hand, for autoregressive-based methods, the process is repeated $|\mathbf{y}|$ times: a) one $\mask$ token is added to the input list $\mathbf{x}$, b) a $1 \times M$ probability distribution of the current masked token is obtained, and c) the input list $\mathbf{x}$ is updated by appending the item with the highest probability.
    \item Result deduplication. The desired outcome is to select the item with the highest probability for each position. However, it may result in duplicates. To resolve this, at each position, we select the item with the highest probability and not present in the previous positions. This leads to a target list of length $|\mathbf{y}|$ with no repeated items.
\end{enumerate}

\subsection{Other Training Details}

We implement Caser\footnote{\url{https://github.com/graytowne/caser\_pytorch}}, SASRec\footnote{\url{https://github.com/kang205/SASRec}}, GRU4Rec\footnote{\url{https://github.com/hidasib/GRU4Rec}}, and CAR\footnote{\url{https://github.com/heyunh2015/ListContinuation\_WSDM2020}} based on the original implementations. We use the Transformers~\cite{Transformers} library to implement BERT4Rec and our FANS model in PyTorch, and the source code and data are available at: \url{https://github.com/Jyonn/FANS}. All the methods were trained with NVIDIA GeForce RTX 3090 with 24GB memory. The results are averaged over 5 runs and p-values are smaller than 0.01. The maximum training epoch is set to 200, and the patience of early stop is set to 3.

\section{Inference Time Analysis}

\autoref{fig:per-vc} illustrates the consumed time ratio of the Transformer network and the vanilla classifier in the inference stage of a non-autoregressive model (e.g., \nagbert mentioned in \autoref{sec:ex} ). It can be seen that even with many Transformer layers, the vanilla classifier still dominates the total inference time. To alleviate this problem, we propose to leverage item category information and design a two-stage classifier to hierarchically decode latent representations. As shown in \autoref{fig:per-tsc}, applying the two-stage classifier can greatly reduce the inference time.

\input{plots/per}

%% file: plots/per.tex
\begin{figure}[!h]
    \centering
    \resizebox{1.0\linewidth}{!}{
    \begin{tabular}{cc}
    \begin{subfigure}{0.25\textwidth}
        \resizebox{1.0\linewidth}{!}{
            \includegraphics{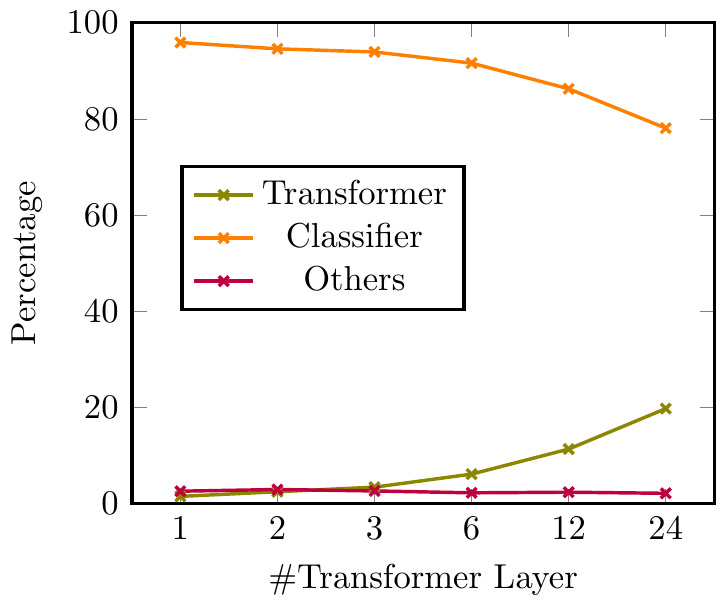}
        }
        \caption{\label{fig:per-vc}Vanilla classifier}
    \end{subfigure} & 
    \begin{subfigure}{0.25\textwidth}
        \resizebox{1.0\linewidth}{!}{
            \includegraphics{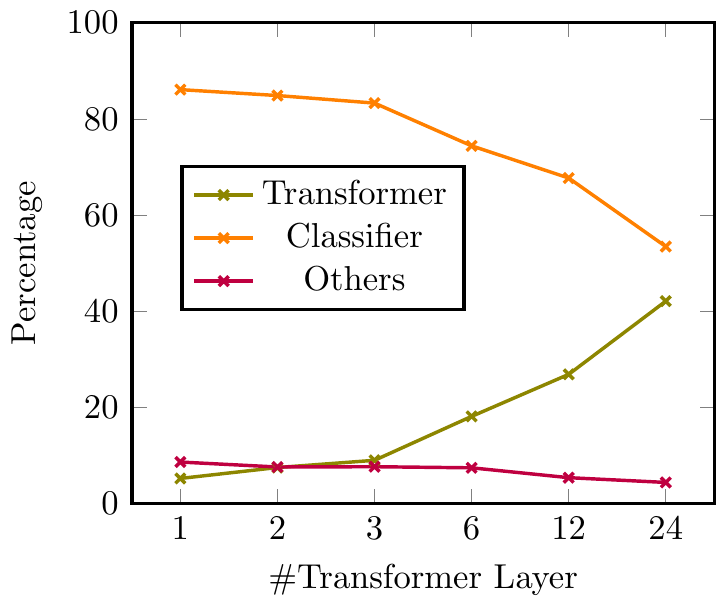}
        }
        \caption{\label{fig:per-tsc}Two-stage classifier}
    \end{subfigure}
    \end{tabular}
    }
    
    \caption{\label{fig:per} Percentage of inference time used by different modules in a non-autoregressive model (e.g., \nagbert). The left one uses the vanilla classifier as in BERT, and the right one uses our proposed two-stage classifier. Specifically, the vanilla classifier calculates probability distribution over the global item vocabulary, while our proposed two-stage classifier first predicts item category and then calculates probability distribution over the local category-specific item vocabulary. The time spent by the classifiers (in orange) includes the sorting operation to obtain top-$K$ items.}
    \Description{Percentage of inference time used by different modules in a non-autoregressive model. The left one uses the vanilla classifier as in BERT, and the right one uses our proposed two-stage classifier. Specifically, the vanilla classifier calculates probability distribution over the global item vocabulary, while our proposed two-stage classifier first predicts item category and then calculates probability distribution over the local category-specific item vocabulary. The time spent by the classifiers (in orange) includes the sorting operation to obtain top-$K$ items.}
\end{figure}